\documentclass{sigchi-ext}
\usepackage[T1]{fontenc}
\usepackage{textcomp}
\usepackage[scaled=.92]{helvet} % for proper fonts
\usepackage{graphicx} % for EPS use the graphics package instead
\usepackage{balance}  % for useful for balancing the last columns
\usepackage{booktabs} % for pretty table rules
\usepackage{ccicons}  % for Creative Commons citation icons
\usepackage{ragged2e} % for tighter hyphenation

\copyrightinfo{ }

\def\plaintitle{Practical Challenges in Indoor Mobile Recommendation}

\def\emptyauthor{}
\def\plainkeywords{Recommendation systems; mobile recommendation; indoor recommendation; systematic review.}
\title{Practical Challenges in\\Indoor Mobile Recommendation}

\numberofauthors{2}
\author{%
  \alignauthor{%
    \textbf{Leandro Marega F. Otani}\\
    \affaddr{Faculdade de Inform\'atica e Administra\c{c}\~ao Paulista (FIAP)}
    \affaddr{S\~ao Paulo, SP, Brazil}\\
    \email{leandro.otani1@gmail.com}\\
    }
    \alignauthor{%
    \textbf{Vagner Figueredo de Santana}\\
    \affaddr{IBM Research, IBM}\\
    \affaddr{S\~{a}o Paulo, SP, Brazil}\\
    \email{vagsant@br.ibm.com}}\\
}
\definecolor{linkColor}{RGB}{6,125,233}
\hypersetup{%
  pdftitle={\plaintitle},
  pdfauthor={\emptyauthor},
  pdfkeywords={\plainkeywords},
  bookmarksnumbered,
  pdfstartview={FitH},
  colorlinks,
  citecolor=black,
  filecolor=black,
  linkcolor=black,
  urlcolor=linkColor,
  breaklinks=true,
}

\begin{document}

\maketitle

% Uncomment to disable hyphenation (not recommended)
% https://twitter.com/anjirokhan/status/546046683331973120
\RaggedRight{} 

%TODO: review the use of bracketed references as nouns
% instead of [12] says that.... use Otani [12] says that...

% In cases of multiple references you can use
% Authors using this approach advocate. ... [12, 23, 34, 35, 46, 47,48,49,50]

% Do not change the page size or page settings.
\begin{abstract}
Recommendation systems are present in multiple contexts as e-commerce, websites, and media streaming services. As scenarios get more complex, techniques and tools have to consider a number of variables. When recommending services/products to mobile users while they are in indoor environments next to the object of the recommendation, variables as location, interests, route, and interaction logs also need to be taken into account. In this context, this work discusses the practical challenges inherent to the context of indoor mobile recommendation (e.g., mall, parking lot, museum, among others) grounded on a case and a systematic review. With the presented results, one expects to support practitioners in the task of defining the proper approach, technology, and notification method when recommending services/products to mobile users in indoor environments.
\end{abstract}


%\category{H.5.m}{Information interfaces and presentation (e.g.,
 % HCI)}{Miscellaneous}
  %\category{See}{\url{http://acm.org/about/class/1998/}}
  %{for full list of ACM classifiers. This section is required.}

%TODO: review the use of bracketed references as nouns
% instead of [12] says that.... use Otani [12] says that...

% In cases of multiple references you can use
% Authors using this approach advocate. ... [12, 23, 34, 35, 46, 47,48,49,50]

\newpage

\section{1. Introduction}

Recommendation Systems are tools and techniques that can be used to suggest relevant content to a user in order to support decision-making processes~\cite{recommender_handbook,internet_rec_sys}. Popular examples of such systems are Netflix~\cite{netflix_home} and Amazon~\cite{amazon_home}, where users can discover new and relevant content based on their preferences and activity.

Beyond the concept of Recommendation System, it is common to find diverse classifications in the literature to specify the data source considered~\cite{hybrid_recommender_survey}. The three main approaches are the following:
\begin{itemize}
	\item \textbf{Collaborative filtering:} A domain-independent and profile-based technique that builds a database of user profiles in order to provide recommendation for users with similar preferences. The main limitation of this technique is that it depends on registered preferences.
	\item \textbf{Content-based recommendation:} A technique in which the recommendation occurs based on user previous content accesses, taking into account item similarities. The similarity is based on items' common features. The main limitation of this technique is that it depends on in-depth knowledge and rich description of features related to recommendable items.
    \item \textbf{Hybrid filtering:} Combines collaborative and content-based approaches in order to improve recommendations and to reduce disadvantages of both approaches~\cite{cont_collab_recomm,rec_sys_principles,content_based_reco}.
\end{itemize}

In addition to the three approaches presented, there are works towards indoor mobile recommendation systems~\cite{mobile_rec_sys} and location-based recommendation~\cite{loc_based_recom}, in which the main objective is to provide a personalized experience by using user's location.

In this context, this work aims at discussing the practical challenges related to the main techniques used in indoor mobile recommendation systems. The questions driving this discussion are the following: (1) What is the most used recommendation approach? (2) What is the most used method to define the right moment/location to provide a recommendation? (3) What is the most used technology to support the indoor recommendation? (4) Do the studies count on user tests? If yes, how many participants were involved?

This paper is organized as follows: section 2 details the review methodology, section 3 presents the results, and section 4 discusses the main techniques for indoor mobile location and connects with practical challenges of a case.

\section{2. Methodology}
First, in order to identify the main techniques and answer the questions driving the discussion proposed in this work, we followed the methodology presented by Kitchenham~\cite{systematic_review_guide} to perform a Systematic Review. % The details of the reviewing process are defined next.

Starting from the theme, the keywords that describe the subject were selected and a query string was elaborated with corresponding search terms. The chosen keywords are: "indoor", "mobile" and "recommendation"; Table~\ref{tab:table1} presents keywords and search terms defined.

\begin{table}
  \centering
  \begin{tabular}{l p{3cm}}
    % \toprule
    {\small\textit{Keyword}}
    & {\small \textit{Search term}} \\
    \midrule   
	Indoor & indoor \newline inside building \newline inside room \\
    Mobile & mobile \newline smartphone \newline cellphone \newline smart phone \\
    Recommendation & recommendation \newline recommender \newline notification \newline advertisement \\
    % \bottomrule
  \end{tabular}
  \caption{Keywords and search terms used.}~\label{tab:table1}
\end{table}

\subsection{2.1. Resources and Criteria}
For the search stage, three digital libraries were chosen: ACM, IEEE, and SpringerLink. 
The inclusion and exclusion criteria are defined next. 
The inclusion criteria consider materials: (1) Published between January 2010 (popularization of smartphones with geolocation support) and January 2018 (the time that this work took place); (2) Written in English; (3) Published on a periodic or conference, and; (4) Addressing the context of indoor mobile recommendation. 
The exclusion criteria considers materials: (1) Published in a time range incompatible with the one defined; (2) Written in a language different from English; (3) Published as books, reports, white papers, conference program,
\footnote{To ensure this criterion for SpringerLink library, the query string used the term "NOT ABSTRACTS" to reduce the noise in the initial search result, removing works published only as abstracts, as usually occurs in conference programs.}
or any other material other than the ones defined in the inclusion criteria, and; (4) Are not related to the previously defined keywords nor addressing the desired context.

\begin{marginfigure}[-35pc]
  \begin{minipage}{\marginparwidth}
    \centering
    \includegraphics[width=0.9\marginparwidth]{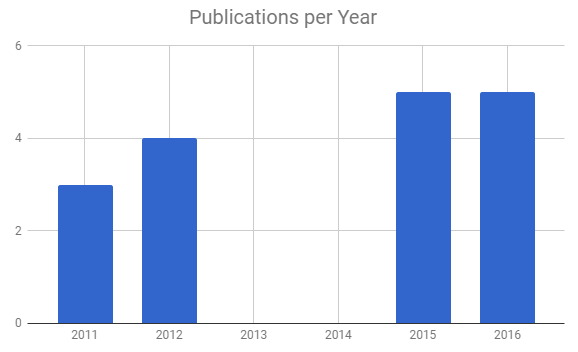}
    \caption{Number of publications per year, excluding 2017.}~\label{fig:figure1}
  \end{minipage}
\end{marginfigure}

\subsection{2.2. Extraction and Annotation Process}
With the search terms and criteria defined, the extraction and annotation process took place. The resulting sets of papers extracted from each library are presented in Table~\ref{tab:table2}. The filtering order follows the criteria order, in which the first one is related to time range, the second one to the material language, third one to the publication type, and the fourth one to the subject.

\begin{table}[!htp]
  \centering
  \begin{tabular}{l r r r r r}
    % \toprule
    & & \multicolumn{4}{c}{\small{\textbf{Filters}}} \\
    \cmidrule(r){3-6}
    {\small\textit{Library}}
    & {\small \textit{Results}}
    & {\small \textit{1st}} 
    & {\small \textit{2nd}} 
    & {\small \textit{3rd}} 
    & {\small \textit{Final}} \\
    \midrule
    ACM & 69 & 40 & 40 & 35 & 2 \\
    IEEE & 136 & 92 & 92 & 89 & 16 \\
    SpringerLink & 1364 & 282 & 280 & 4 & 0\\
    %ACM & 59 & 30 & 30 & 27 & 2 \\
    %IEEE & 128 & 84 & 84 & 83 & 15 \\
    %SpringerLink & 1334 & 252 & 250 & 0 & 0\\
    % \bottomrule
  \end{tabular}
  \caption{Library database search results at each filtering stage.}~\label{tab:table2}
\end{table}

\section{3. Review Results}
The filtering process resulted in 18 papers, in which a characterization was applied in order to provide an overview of contributions by year, author, country and objective (academic research, industry research, or both). Regarding the year of publication, there is a returning interest of the community through the years of 2015 and 2016, after 2 years of low activity in the subject (2013 and 2014).

The contributions came from 14 different countries, whereas China and Japan lead in number of publications (5 and 4, respectively) and authors (28 and 12, respectively). In this set of papers, there was no author with more than one publication.

With respect to objectives, 72.22\% of the contributions came from academic research whereas industry-only research and joint collaborations represented 11.11\% and 16.66\% of the papers, respectively.

\section{4. Secondary Results}

Once the filtered dataset was obtained, the following questions were defined in order to guide the discussion on practical challenges faced by the ones working on indoor mobile recommendation:
\begin{enumerate}
  \item What is the most used recommendation approach?
  \item What is the most used method to define the right moment/location to provide a recommendation?
  \item What is the most used technology to support the indoor recommendation?
  \item Do the studies count on user tests? If so, how many participants were involved?
\end{enumerate} 

In the next subsections, we summarize and discuss the results obtained for each of these questions.

\subsection{3.1. What is the most used recommendation approach?}

Regarding the recommendation technique used, the papers were classified according to the main recommendation approaches: (i) Content-based recommendation; (ii) Collaborative filtering; (iii) Hybrid filtering; (iv) Other. 

\begin{marginfigure}[-35pc]
  \begin{minipage}{\marginparwidth}
    \centering
    \includegraphics[width=0.9\marginparwidth]{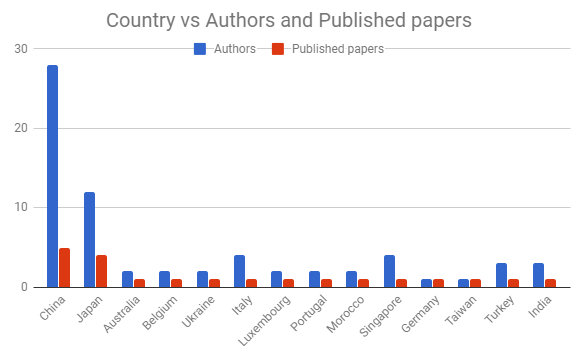}
    \caption{Number of authors and published works on each country, over the period of 2010 - 2017.}~\label{fig:figure2}
  \end{minipage}
\end{marginfigure}

The most used technique is the \textit{content-based}, present in 8 papers~\cite{mambo_advert, magnetic_based_prox, ibeacon_museum_hall, dev_astronomy, using_selfhst, pers_recsys, flexible_campus, aal_sols} (44.44\%). The \textit{collaborative filtering} and \textit{hybrid filtering} categories are present in 4~\cite{adnext, encounter_meet_alg, parkgauge, shopassist} (22,2\%) and 2~\cite{ibeacon_route_guidance,ionavi} (11,11\%) papers, respectively. Four papers were classified as \textit{other}. One of them mentioned the usage of linear matching, which involves the analysis of different items' characteristics (in this case images) in order to identify similarities~\cite{linear_matching}. The authors advocate the use of linear matching when lacking historical data. The other three~\cite{determining_loc, ble_geomarket, smartshoppe} focused on recommendations based on user location, being strictly dependent on proximity detection, such as passing-by events.

\begin{marginfigure}[-35pc]
  \begin{minipage}{\marginparwidth}
    \centering
    \includegraphics[width=0.9\marginparwidth]{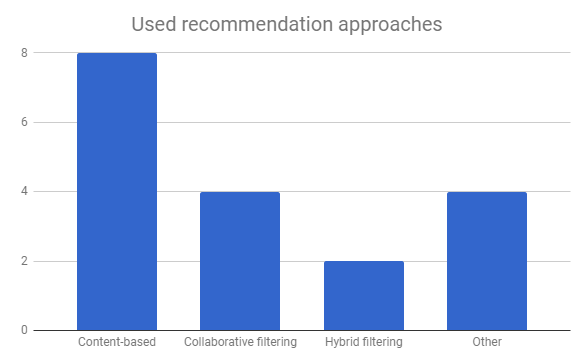}
    \caption{Recommendation approaches used in the set of papers studied.}~\label{fig:figure3}
  \end{minipage}
\end{marginfigure}

\subsection{3.2. What is the most used method to define the right moment or location to provide a recommendation?}

Most of the articles did not detail the algorithm used to provide the recommendation. However, as presented in the previously, \textit{content-based} approach was the most applied approach. In addition, there was a proposal~\cite{bayesian_network_collab} of using Bayesian Networks in \textit{collaborative filtering} and, also, some specific algorithms, such as EncounterMeet~\cite{encounter_meet_alg}, which is also based on \textit{collaborative filtering}. Regarding the time and place for triggering a recommendation, most of the papers (82,35\%) were based on position triangulation algorithms to provide location-specific recommendation. 
%The techniques were: Proximity detection from a single device, % TODO: Complementar

Regarding delivery methods, the following alternatives were identified:
\begin{enumerate}
	\item \textbf{Push notification:} Choice of 50\% of the articles~\cite{adnext, magnetic_based_prox,ibeacon_museum_hall,ibeacon_route_guidance,dev_astronomy,using_selfhst,ble_geomarket,shopassist,smartshoppe}, it can be defined as a technique that delivers messages to user's smartphone relating to an existing application, even when that application is not running~\cite{push_notifications}.
   \item \textbf{Interaction-dependent recommendation:} Choice of 22.22\% of the articles~\cite{linear_matching, parkgauge,aal_sols,ionavi}, it considers recommendations presented as user interacts with the mobile application.
   \item \textbf{Hybrid method:} Chosen of 27.77\% of the articles~\cite{det_location, encounter_meet_alg, mambo_advert, pers_recsys, flexible_campus}, it considers articles that used both interaction-dependent and push notification methods.
\end{enumerate}

\subsection{3.3. What is the most used technology to support the indoor recommendation?}

From the technological perspective, the two main technologies used were Beacon, present in 8~\cite{ibeacon_museum_hall,ibeacon_route_guidance,dev_astronomy,pers_recsys,flexible_campus,ble_geomarket,shopassist,smartshoppe} papers (44.44\%), and wi-fi, present in 4~\cite{bayesian_network_collab,determining_loc,encounter_meet_alg,ionavi} papers (22.22\%).

% \begin{marginfigure}[-35pc]
%   \begin{minipage}{\marginparwidth}
%     \centering
%     \includegraphics[width=0.9\marginparwidth]{figures/technology-vs-papers}
%     \caption{Technologies used in the set of papers studied.}~\label{fig:figure4}
%   \end{minipage}
% \end{marginfigure}

The beacon technology was initially introduced by Apple~\cite{apple_ibeacon} in 2013, being designed to advertise through small stationary hardware packages that included unique identification attributes and proximity to mobile devices, powered by the Bluetooth 4.0 Low Energy (BLE) architecture~\cite{bluetooth_le}. It is also worth noting that approaches using wi-fi were published mainly in 2011 (3 papers), whereas beacons have been cited in 2015 (4 papers) and 2016 (3 papers). No publication was found mentioning beacon in 2017. This may be related to the timespan this review took place and the time required to index publications in the digital libraries used.

The other proposed technologies, with one citation each, included: Radio-Frequency Identification (RFID), 2D visual markers, magnetic induction, and augmented reality, where the first three ones are all proximity-based technologies. There were also 2 papers that did not present a specific technology, focusing primarily on recommendation algorithm or a general method/architecture.

In an analysis involving performance and benefits of these technologies, Jiang~\cite{magnetic_based_prox} verified that: (i) The BLE has large gray regions (regions where detections happen in an unpredictable manner) and become inconsistent over time; (ii) The RFID has sharp boundaries (balance between gray zones and well covered zones), but suffers heavily from attenuation from obstacles such as the human body; (iii) The magnetic induction approach has sharp, consistent boundaries. Regarding beacon technologies, a few positive aspects were also pointed out, such as lower acquisition cost, low energy consumption, and wide compatibility~\cite{ibeacon_museum_hall,ibeacon_route_guidance}.

\subsection{3.4. Do the studies count on user tests? If so, how many participants were involved?}

From the 18 papers selected, 8 papers (44.44\%) reported some sort of user test. From those 8 papers, 2~\cite{ibeacon_route_guidance,flexible_campus} did not inform how many users participated in the test, thus, they are not considered for this analysis. In the 6 papers with user tests, the mean number of participants is 19.5 (standard deviation of 20.52). The contexts of these 6 studies involved mobile social networking, conference event, garage parking, subway to outdoor navigation, proximity-based recommendation, and museum.

% Comentei este trecho para termos mais espaço para a discussão (@vagsant)
The considered experiments are contextualized bellow:

\begin{enumerate}
	\item Mobile social networking: 10 participants were asked to evaluate up to 10 friend recommendations based in a common friends algorithm and another 10 friend recommendations, based on an algorithm being proposed. The evaluation resulted in a good acceptance (44.6\% vs. 32.1\%) of the proposed algorithm~\cite{encounter_meet_alg}.

	\item Conference event: 59 participants received recommendations such as relevant talks, posters and people to meet. Through the evaluation, the author concluded that the recommendation had a conversion rate of 12\%, pointed as good by the author once analyzed as a serendipity rate~\cite{pers_recsys}.

	\item Garage parking: 6 drivers performed parking maneuvers in different shopping malls in order to test slot a occupancy detection algorithm. The author pointed that results have shown low error in parking identification~\cite{parkgauge}.

	\item Subway to outdoor navigation: 20 participants were asked to define destinations that started from three different subway stations. The results have shown up to 100\% route recommendation performance~\cite{ionavi}.

	\item Proximity based recommendations: 3 participants that tested the proximity based advertising in passing-by and staying scenarios. According to the authors, the result was a sharp and consistent proximity detection~\cite{magnetic_based_prox}.

	\item Exposition recommendation: 19 participants could freely navigate on the 5th floor of a museum in order to test automated proximity based recommendations effectiveness in comparison with a recommendation based on screen interaction. In a post-experiment evaluation, 8 participants found the automated recommendation effective, 5 did not, 4 did not answer, and 2 found both methods effective\cite{ibeacon_museum_hall}.

\end{enumerate}

\section{4. Discussion}
The following discussion is divided into three parts: (1) The systematic review, in which we present a summary of our findings along with relevant considerations and possible implications; (2) The PURPLE mobile application, which discusses a practical case involving technology trials and pilot user tests; (3) Final considerations and future directions.

\subsection{5.1. The systematic review}
%TODO: Discuss the results and the impacts
As shown previously, mobile recommendation in indoor environments is gaining attention in recent years. This fact might be related to advances in smartphone hardware~\cite{cpu_rise_power} and the popularization of technologies such as BLE, a key component of Internet of Things (IoT)~\cite{internet_of_things} devices, including Beacon and similar sensors.

% TODO: Discuss how the results can be applied in recommender systems
Regarding the recommendation approach, there is a trend towards the use of content-based approach, associated with daily activities encompassing mobility and social events. In those contexts, few of the identified studies presented user experiments reporting good user acceptance towards mobile recommendation~\cite{ibeacon_museum_hall,pers_recsys,encounter_meet_alg}. It is worth highlighting that acceptance has a significant role in all of those contexts, especially in mobile marketing scenarios. In this sense, literature also points out that proper definition of moment, location, and method of recommendation are key aspects for user engagement and satisfaction~\cite{mini_q_mob_marketing_value,mini_q_consum_attitudes,mini_q_cross_market}.

Another relevant issue is the \textit{cold start} problem~\cite{cold_start_problem}, addressed by Kaya~\cite{linear_matching} with the linear matching algorithm proposal, but barely discussed by other authors. This was somewhat expected, given that solutions found where mainly focused on high-level aspects such as efficiency of delivery methods and architecture of the solution.

\subsection{5.2. The PURPLE mobile application}

In alignment with the previous aspects present in the literature, it is also relevant to share results obtained after trials of providing indoor mobile recommendation, user feedback, and technical viability. These results are related to a case involving the PURPLE mobile application, a platform focused on personalized content delivery that leverages the use of Beacons and smartphones to recommend relevant content based on user's location, interests, and analysis of indoor behavior.

This case is situated in the Brazilian context and was first idealized to be used in malls. However, the first two pilots were in the academic conferences context; taking place during the XV and XVI editions of the \textit{Brazilian Symposium on Human Factors in Computer Systems} (IHC 2016\footnote{http://ihc2016.mybluemix.net/} and IHC 2017\footnote{http://ihc2017.ihcbrasil.com/}). The first pilot considered the technological challenges, while the second pilot aimed at exploring notification mechanism.

The pilots involved a smartphone application with agenda, local attractions recommendations, presentations/sessions reminders, and event announcements, all of them considering users' indoor location. Those pilots had a special focus on the technical challenges and the viability of the delivery method in the country's context. In sum, the challenges faced were:
to characterize users in order to determine available gadgets and users' interests; to identify factors that could influence the adoption of the mobile application during the whole event; to develop a solution compatible with a variety of devices ranging from low budget Androids~\cite{android}/old iPhones~\cite{iphone} to high-end devices; to develop a battery efficient solution that constantly uses GPS/BLE; to identify infrastructure limitations of such events.

As a starting point, the characterization of the community did play an important role in defining relevant features, such as local attractions recommendation and places to eat, along with most used device brand and Operational System. As a solution, the characterization method presented in~\cite{characterization_community} was used to initially describe the community, taking into account the user footprint information available through the use of Google Analytics~\cite{google_analytics}, configured at the IHC 2016 website.

Regarding adoption, a few key aspects could be pointed out as influential factors: existence of printed alternatives to application functionality, such as tourism guides, agenda, and announcements; official support by event organization; advertisement. For open experiments such as the one presented, natural degradation of user adherence is also expected, once users need to be interested not only in installing and opening the application for the first time, but actually be constantly engaged during upcoming days.

Considered those key aspects, defining proper technology to make the application widely available through different devices in a short period of time is valuable in order to conquer an early partnership with the event organization and, consequently, proper advertising. From the various implementation alternatives, the following hybrid mobile development frameworks are interesting choices: 
\begin{itemize}
\item Cordova~\cite{cordova}, the chosen framework for IHC 2016 mobile application, it is a web-based, \textit{code-once and run everywhere} solution that is widely supported by its community and has auxiliary functionality add-ons, but lacks performance and is susceptible to bugs/customizations present in the device's browser.
\item React Native~\cite{react_native}, the chosen framework for IHC 2017 mobile application, counts on a better development auxiliary tool set, improved application performance, and a standardized behavior across all the devices, in exchange for a higher implementation complexity.
\end{itemize}

In the first pilot (IHC 2016), where we collected feedback from 62 users, no complaints were registered regarding the smartphone battery consumption. This was achieved due to precautions taken when defining the timespan for each smartphone-server package exchange, especially regarding GPS information, as it tends to drain more power than the BLE communication with the Beacons.

Finally, as of infrastructure limitations, the main concern was regarding wi-fi and smartphone connection, in which the device storage was used as a buffer during insufficient connection periods. Here, the lightweight network communication protocol used was MQTT~\cite{mqtt}. 

% TODO: Present what you could change in Purple considering these results
% Personally, not really sure if there is something i would change, as the results showed that our directions are compatible with the research tendency, and fill the gap related to the lack of experiments and studies that are focused on the USER side of research, much more related to the actual effectiveness of the recommendation other than the technology efficiency or viability of such a paradigm of interaction (flow of interaction events and modals, architecture etc)

% TODO: Discuss how the results support developers, etc.
\subsection{5.3. Final Considerations}
With all considerations presented and the goal of supporting developers, human factors practitioners, and marketing professionals in the task of defining a practical indoor mobile recommendation approach, technology, and notification method, the follow observations can be made:

\begin{enumerate}
\item Indoor mobile recommendation is a relatively recent field in the sense that much of the technology needed to produce and validate new user interaction paradigms is becoming available at this very moment. Thus, although it is possible to elect a preferred recommendation approach and delivery method, the current discussion is much more focused on overall software solution than effectively recommendation in terms of user interests and preferences.
\item Regarding technology, a trend from older technologies (e.g., Wi-Fi) to the current era of BLE sensors is in place. Hence, the focus is on providing low-cost and effective means of triangulating user position in order to provide location-based services.
\item Real world scenarios must be taken in to account, focusing on user interaction in order to define the most convenient approaches of recommendation considering user satisfaction, the overall solution scalability, and notification conversion rate.
\item Characterization of user community was proven to be an important subject in the pilots described as it could influence in determining proper technology and engagement strategies to keep constant adherence through the event.
\end{enumerate} 

Once having those findings in mind, developers can take relevant directions towards a time efficient development stack, considering technology trends and main recommendation approaches along with (dis)advantages in their contexts. At the same time, it is expected that human factors and marketing professionals to find an opportunity to connect specific knowledge of their respective domains with the presented results, thus measuring cost, risks, and opportunities of such advertising approach and, by doing so, collaborate towards filling the gaps between technology and user needs.

\section{6. Conclusions}

This paper presented a systematic review on indoor mobile recommendation. The research questions that guided the review pointed that: (1) The most used approach is content-based recommendation; (2) There wasn't a clear definition regarding algorithms used to define the right time/location to provide the recommendation; (3) The technology most used in indoor mobile recommendation systems is beacon and wi-fi -based methods; (4) Few papers count on user tests to validate results in the real environment, highlighting the need for partnerships while validating/evaluating indoor mobile recommendation systems.

This paper contributed with a systematic review on indoor mobile recommendation systems showing that there is an increasing interest in the theme and that the triangulation techniques using beacon devices are promising directions to be followed by developers, human factors, and marketing professionals working on indoor mobile recommendation systems.

Future work involves developing and validating an indoor mobile recommendation system following the trends and outcomes presented in this systematic review, avoiding the shortcomings presented and using the most of the available technology from both enterprise and user sides.

\bibliographystyle{SIGCHI-Reference-Format}
\bibliography{sample}

\end{document}